\documentclass[3p,times,twocolumn]{elsarticle}

\usepackage{ecrc}

\volume{00}

\firstpage{1}

\journalname{Nuclear Physics B Proceedings Supplement}

\runauth{}

\jid{nuphbp}

\jnltitlelogo{Nuclear Physics B Proceedings Supplement}

\usepackage{amssymb}

\usepackage[figuresright]{rotating}

\begin{document}

\begin{frontmatter}



\dochead{}

\title{Cosmic ray propagation and interactions in the Galaxy}


\author{V.N.Zirakashvili}

\address{Puskov Institute of Terrestrial Magnetism, Ionosphere and Radiowave
Propagation, Russian Academy of Sciences, 142190 Troitsk Moscow, Russia}

\begin{abstract}
Cosmic ray propagation in the Galaxy is shortly reviewed. In particular we
 consider the self-consistent models of CR propagation. In these models CR streaming instability
 driven by CR anisotropy results in the Alfv\'enic turbulence which in turn determines the
 scattering and diffusion of particles.

\end{abstract}


\end{frontmatter}


\section{Introduction}
\label{}

Diffusion model of Ginzburg and Syrovatsky \cite{ginzburg69} was one of the first physically
 justified models of cosmic ray (CR) propagation. According to this model CR sources are supernova remnants
  (SNRs) which  are situated in the
 Galactic disk. CR particles perform wandering in the tangled magnetic fields of the Galaxy.
The propagation region of CRs is not limited by the Galactic disk but also contains some region
 above and below the Galactic disk - a so called Galactic halo.

Although CR diffusion was introduced  phenomenologically it obtained later the theoretical
 basis \cite{berezinsky90}. Diffusive shock acceleration (DSA) \cite{krymsky77, bell78}
in supernova remnants is considered now
 as a principle mechanism of CR production in the Galaxy. Its main predictions are in accordance
 with modern gamma-ray observations of supernova remnants \cite{rieger13}.

CR confinement time in the Galaxy can be estimated using CR secondaries.  CRs contain a significant
 amount of nuclei that are not abundant in nature. They appear after nuclear
 fragmentation of  primary CRs in the interstellar medium. The measured Boron to Carbon CR ratio is
 shown in Fig.1 \cite{obermeier12}. It is important that the ratio drops when the energy increases.
This means that the
 residence time in the Galactic disk $t_{res}(E)$ is lower for higher energies.
It is convenient to
 use a so called  grammage $\Lambda (E)=v\rho t_{res}$ that is the mean amount of matter
 transversed
 by CR particles. Here $v$ is the speed of the particles and $\rho $ is the mean gas density
in the Galactic disk. The measured secondary to primary
ratio can be used to estimate the grammage $\Lambda (E)$. This gives $\Lambda (E)\propto E^{-\mu }$ at energies
 higher than several GeV per nucleon with
 the index $\mu $ being between 0.3 and 0.6.

In the pure diffusion model the grammage and CR diffusion coefficient are related as
$\Lambda (E)\propto D^{-1}$, so we expect that CR diffusion coefficient increases with energy as
$D\propto E^{\mu }$. The situation is more complicated in the models which take into account
other processes like reacceleration, advection etc.

\section{Cosmic ray diffusion in Galactic magnetic fields. }

CR diffusion is determined by magnetic inhomogeneities.
The scattering of
 particles occurs via interaction with random magnetic fields $\delta B$ with the scales comparable with
 the gyroradius of particles. The scattering frequency $\nu $ can be estimated as

\begin{equation}
\nu \sim \Omega \frac {\delta B^2}{B^2}
\end{equation}
Here $B$ is the regular magnetic field and $\Omega =qBv/pc$ is the gyrofrequency of particles with
 the electric charge $q$ and momentum $p$.
The diffusion coefficient along the regular magnetic field $D_{\parallel }$ is given by the relation
$D_{\parallel }=v^2/3\nu $.

According to modern theories the  MHD turbulence have two main components: anisotropic
quasi-Alfv\'enic incompressible fluctuations with $k^{-5/3}$ spectrum and the  isotropic
 magnetosonic waves with the spectrum $k^{-3/2}$ \cite{brandenburg13}.

The quasi-Alfv\'enic
 magnetic inhomogeneities are elongated along the local magnetic field, so when their length is
 of the order of particle gyroradius the corresponding perpendicular scale is small. That is why
 the scattering by the quasi-Alfv\'enic component is not effective \cite{yan04}.

The second isotropic magnetosonic component
 is good enough for scattering. The corresponding energy dependence of diffusion coefficient
 $D_{\parallel}\sim vp^{1/2}$ is in good agreement with measured secondary to primary ratios
 of Galactic CRs. This possibility is considered as a good physical
solution for the propagation problem \cite{strong07}. 

However the magnetosonic component exists only when MHD approximation of interstellar
turbulence is used. Magnetosonic waves are damped via the linear Landau damping \cite{ginzburg61} in the more
 justified plasma description of interstellar turbulence. This damping prevents nonlinear
 energy transfer of energy to smaller scales for magnetosonic waves. The Landau damping is
 weaker for waves propagating at small angles relative to the magnetic field.
 However this does not help because the rate of nonlinear transfer of energy to small scales is
 also weaker at these angles. So only strongly oblique magnetosonic waves with their high phase
velocities can avoid the damping and can transfer the energy to
smaller and smaller scales. But
 their obliqueness again will result in the inefficient scattering of CR particles.

We conclude that the main components of interstellar turbulence can not provide the scattering of the
 main part of Galactic CRs with energies below 1 PeV. For higher energies gyroradius of particles
 is above 0.1 pc  and these particles in principle can be scattered by the background turbulence.

\begin{figure}
\begin{center}
\includegraphics[width=8.0cm]{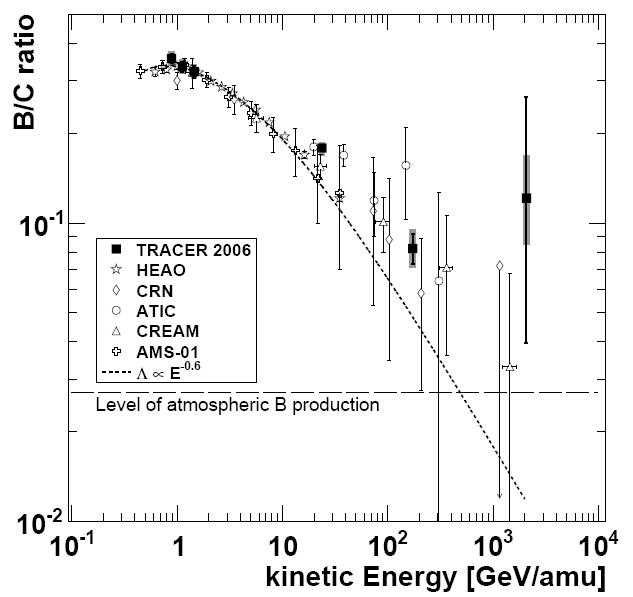}
\caption{Measurements of B/C ratio performed in different
experiments \cite{obermeier12}. }
\end{center}
\end{figure}

\section{Cosmic ray streaming instability and damping of waves}

In this regard another sources of magnetic turbulence should be considered. The best candidate
 is a so called streaming instability driven by anisotropic CR distribution. Its importance for
 CR propagation was recognized many decades ago
 \cite{kulsrud69,wentzel69,kulsrud71,ginzburg73,skilling75}.
The growth rate of unstable Alf\'en waves is given by the 
 equation \cite{lerche67}

\begin{equation}
\Gamma _{CR}\sim \Omega _i\frac {N(r_g>k^{-1})}{n}\left(\frac {u_{cr}}{v_A}-1 \right) .
\end{equation}
Here $\Omega _i$ is the gyrofrequency of thermal ions, $n$ is the plasma number density and
$N(r_g>k^{-1})$ is the number density of CR particles with gyroradii $r_g=pc/qB$ higher than the inverse
 wavenumber $k$. The instability develops when the mean velocity of CR distribution $u_{cr}$ is
higher than the Alfv\'en velocity $v_A$.

CR streaming produces waves with the scale $k^{-1}$ comparable with the gyroradius of particles. The
 particles in turn are scattered by these waves.

Alfv\'en waves produced by GeV particles have the growth rate
$\Gamma \sim 10^{-10}$ s$^{-1}$. So the growth time is only 300
years that is this time is very short in comparison with other
 Galactic time scales.

Some damping mechanisms must be used to prevent the strong growth
of waves. In the warm partially ionized
 regions of the Galactic disk the damping rate of Alfv\'en waves by neutral atoms is given
by \cite{kulsrud71}

\begin{equation}
\Gamma _n=\frac {1}{2}v_{th}\sigma _{ex}n_n
\end{equation}
Here $n_n$ is the number density of neutrals, $v_{th}$ is the
thermal velocity of plasma ions and
 neutral atoms and $\sigma _{ex}$ is the charge exchange cross-section. For the charge exchange
cross-section $\sigma _{ex} \sim 10^{-14}$ cm$^2$ and thermal velocity $v_{th}\sim 10$ km s$^{-1}$
 the damping rate is $\Gamma \sim 10^{-9}$ s$^{-1}$, that is the streaming instability is suppressed in the
 warm interstellar medium. If so one can expect the weak scattering of particles and the large CR diffusion
 coefficient in the Galactic disk.

However neutral atoms absent in the ionized parts of the Galactic
disk and in the Galactic halo where
 CRs propagate. In these regions the existence of the damping of Alfv\'en waves by background
anisotropic MHD turbulence was recognized recently \cite{farmer04,yan04}.
The plasma motions of the background turbulence
 mix the plasma material in the perpendicular to the local magnetic field directions.
As a result the perpendicular wave
 number of the test Alfv\'en waves increases. CR scattering by such waves becomes inefficient
 while the waves eventually absorbed by the background turbulence. The rate of this process for
 oblique waves can be estimated as

\begin{equation}
\Gamma _b\sim \frac {\omega }{(k_{\perp }L)^{1/3}}, \ k_{\parallel }\sim k_{\perp }
\end{equation}
Here $\omega=v_A|k_{\parallel }|$ is the wave frequency,
$k_{\parallel }$ and $k_{\perp }$ are the wavenumbers in the
parallel and perpendicular to the
 local magnetic field direction, $L$ is the scale of the background turbulence. For waves resonant with
 GeV CR particles $k\sim 10^{12}$ cm$^{-1}$ and for the main scale $L=100$ pc we found the damping rate
 $\Gamma _{b}\sim 10^{-9}$ s$^{-1}$. That is the streaming instability of oblique waves is suppressed
in the  ionized regions of the Galactic disk.

However CR streaming produces waves predominantly in the direction of the local magnetic field. For this
 case the damping by background turbulence is weaker and is given by

\begin{equation}
\Gamma _b\sim \frac {\omega }{(k_{\parallel }L)^{1/2}}
\end{equation}

Now the estimate of the damping rate $\Gamma _b\sim 10^{-10}$
s$^{-1}$ is comparable with the expected value for the wave growth of
the streaming instability. For higher energies the damping by the
background turbulence is stronger than the
streaming instability. That is why we conclude that CR
streaming instability can be suppressed in the ionized regions of the
Galactic disk.

This consideration leaves the Galactic halo as the place where CR streaming instability can operate.
It is expected that sources of background turbulence are weaker there. In addition the streaming
instability is faster in the Galactic halo because of the low plasma density (see Eq. (2)).

There exists nonlinear damping of waves in addition to the linear kinds of damping considered above.
 Thermal ions can interact  with moving magnetic mirrors appearing when Alfv\'en waves present in the
 plasma. The gain of the ion energy is accompanied by the corresponding damping of the waves
\cite{lee73,achterberg81,kulsrud82,achterberg86,fedorenko88,zirakashvili00}.
The damping rate can be estimated as

\begin{equation}
\Gamma _{nl}\sim \omega \delta \frac {kW(k)}{B^2/4\pi }
\end{equation}
Here the spectral energy density of waves $W(k)$ is normalized as $\left< \delta B^2\right>=4\pi
 \int dkW(k)$, $\delta $
 is dimensionless parameter of the order of unity. The nonlinear damping rate depends on the
 wave spectrum and can be used to determine the level of the Alfv\'en turbulence generated by
 CR streaming instability.

\section{Galactic wind driven by cosmic rays}

CR influence on the Galaxy is not reduced to generation of Alfv\'enic turbulence. CR energy density is
 comparable with the gas and magnetic energy densities in the Galactic disk. It is expected that
 the propagation region of CRs is significantly broader than the Galactic disk. If so the dynamical
 effects of CRs will be stronger in the Galactic halo where gas density and pressure are lower.
It is possible that CR pressure gradient drives outflow from the Galactic disk and from the Galaxy -
 a so called Galactic wind
\cite{ipavich75,breitschwerdt91,zirakashvili96,everett07}.
The expected geometry of the wind flow is shown in Fig.2. Galactic wind
 flows along the surface $S$. The frozen magnetic field ${\bf B}$ is tangent  to this surface. At large distances
 from the Galaxy the field is almost azimuthal due to rotation of the Galaxy. This azimuthal
 configuration results in the better confinement of high-energy cosmic rays.

\begin{figure}
\begin{center}
\includegraphics[width=7.0cm]{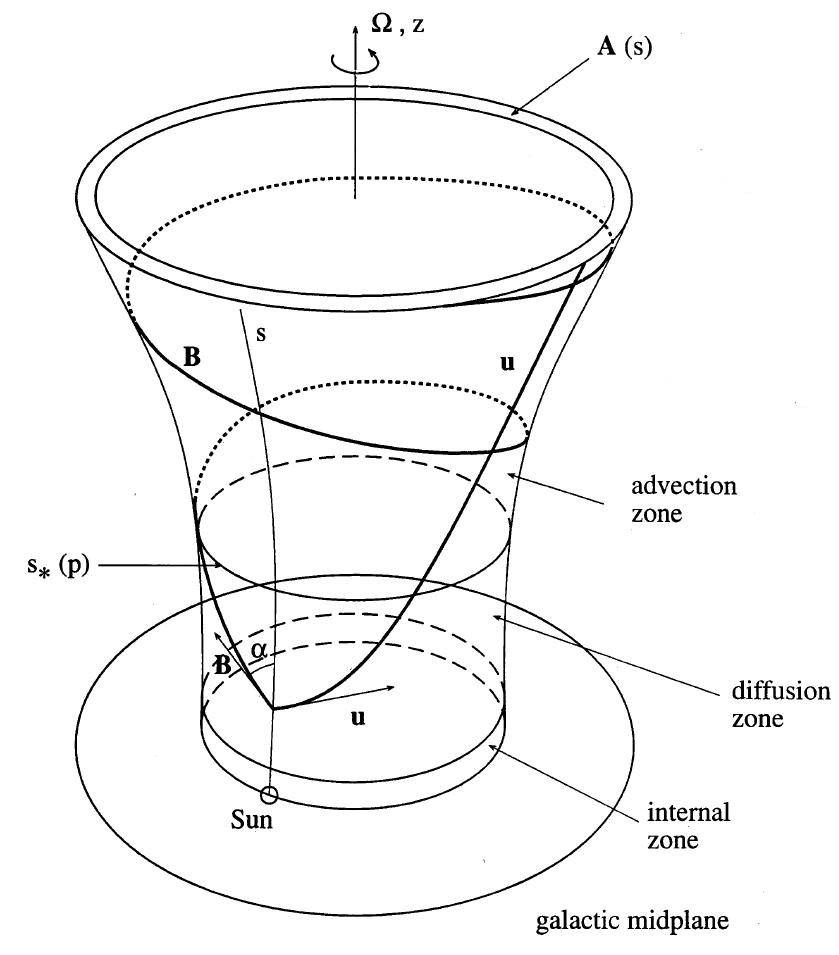}
\caption{Geometry of the flux tube and magnetic field along the
surface $S$. Vectors of magnetic field {\bf B} and gas
velocity {\bf u} are tangent to the surface $S$. $\protect\alpha $
is the angle between the magnetic field and meridional direction. }
\end{center}
\end{figure}

For illustration we show the results of the Galactic wind calculations \cite{zirakashvili02} in Fig.3.
The damping of Alfv\'en waves produced by CR streaming instability results in the strong gas heating
 in the Galactic halo. At low heights this heating is balanced by radiative cooling of the gas. However
 at large heights the gas number density is low and the cooling is not effective. As a result the gas
temperature is of the order of one million degrees at large heights. It is interesting that such a halo
  of the hot gas with a similar density profile is indeed observed now via measurements of OVII line
absorption \cite{miller13}.

\begin{figure}[t]
\begin{center}
\includegraphics[width=8.0cm]{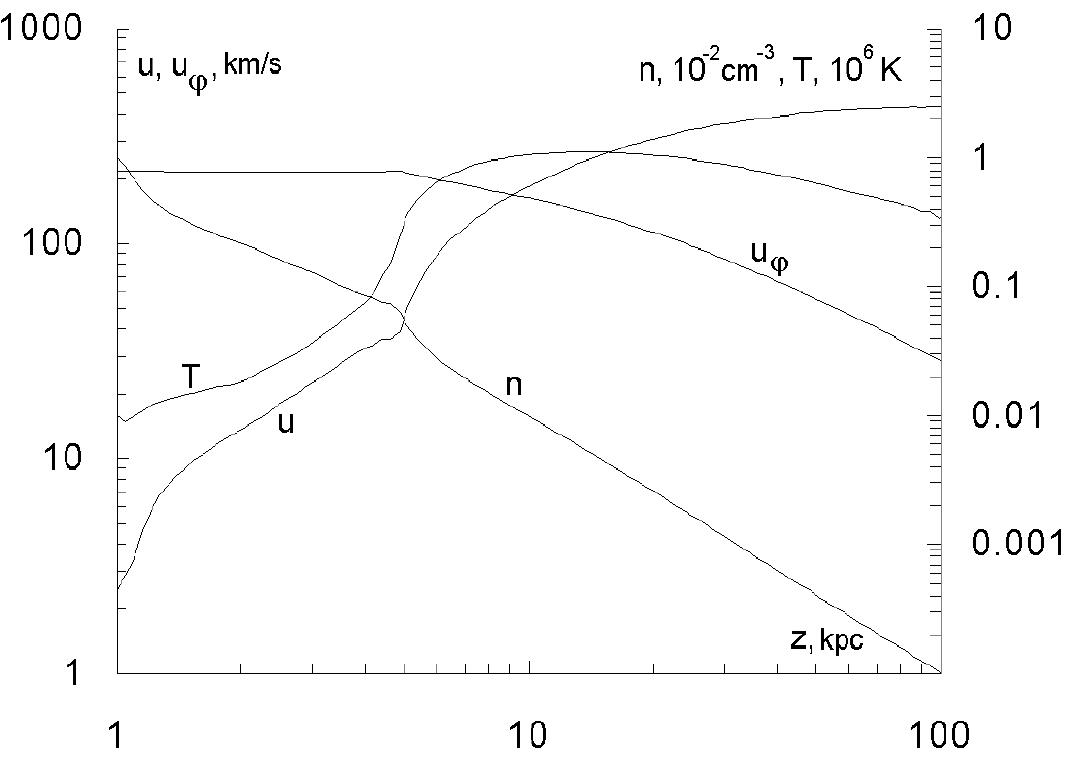}
\caption{
Dependence of meridional and azimuthal components of the gas velocity
 $u$ and
$u_{\phi }$, the gas number density and temperature $T$ on the
height above the Galactic disk $z$. The following parameters were
fixed at 1 kpc above the midplane of the disk: $P_{c0}=5\cdot
10^{-13}$ erg cm$^{3}$, $n _{0}=10^{-2}$ cm$^{-3}$,
$B_{0}=2\cdot 10^{-6}$ G.}
\end{center}
\end{figure}

\section{Self-consistent models of CR propagation}

Steady state CR transport along the surface S is described by the
following equation for isotropic part of CR momentum
 distribution $N(p)$ \cite{ptuskin97}:

\[
\frac 1{A(s)}\frac {\partial }{\partial s}A(s)D_{\parallel }\cos ^2\alpha \frac {\partial N}{\partial s}
-(u+v_a)\frac {\partial N}{\partial s}+
\]
\begin{equation}
\frac {p}{3}\frac {\partial N}{\partial p}
\frac {1}{A(s)}\frac {\partial }{\partial s}A(s)(u+v_a)+2Q(p)\delta (s)=0.
\end{equation}
Here $Q(p)$ is the half of the surface CR source  power and $v_a=\pm v_A\cos \alpha $ is the
meridional component of the Alfv\'en velocity directed away from the Galactic disk.

Let us assume that CR sources in the disk have power-law
dependence on momentum
\begin{equation}
Q(p)=\frac {\varepsilon }{4\pi c(mc)^4}\left( \frac {p}{mc}\right) ^{-\gamma }
H(p_m-p).
\end{equation}
Here $m$ is the proton mass, $p_m$ is the maximum momentum of CR sources, and the parameter
 $\varepsilon $ determines the surface flux of CR energy $F_c$:
\begin{equation}
F_c=\varepsilon \int \limits ^{p_m/mc}_0y^{2-\gamma }dy\left( \sqrt{y^2+1}-1\right).
\end{equation}

We shall assume that CR diffusion is determined by self-excited Alfv\'en waves.
The condition of the local balance for the wave generation and damping $\Gamma _{CR}=\Gamma _{nl}$
 can be used to obtain the spectrum of waves $W(k)$. It in turn determines CR diffusion
coefficient \cite{ptuskin97}:

\[
D_{\parallel }=\frac {8\delta \gamma (\gamma -2)}{9\pi ^2}\frac {c^2}{\Omega _{i}}
\frac {B^2/4\pi }{\varepsilon /c}\left( \frac {p}{mc}\right) ^{\gamma -3}\times
\]
\begin{equation}
A(s)\cos \alpha ,
\end{equation}

The numerical value of $D_{\parallel }$ for $\gamma $ close to 4 is given by expression
\begin{equation}
D_{\parallel }=1.8\cdot 10^{26}\delta \frac {\mbox{ cm}^2}{\mbox{s}}
\left( \frac p{mc}\right) ^{\gamma -3}
B_{\mu G}{\varepsilon }_{-6}^{-1}A(s)\cos \alpha .
\end{equation}
Here we normalize the parameter $\varepsilon $ to its characteristic value
$10^{-6}$ erg cm$^{-2}$ s$^{-1}$.

\subsection{self-consistent diffusion models}

These equations can be used in self-consistent diffusion models of CR
propagation \cite{kulsrud69,wentzel69,kulsrud71,ginzburg73,skilling75,alloisio13}.

Since the observable
 spectrum $N_{obs}(p)\sim Q(p)/D_{\parallel}$ we find that

\begin{equation}
N_{obs}(p)\sim p^{-(2\gamma -3)}
\end{equation}

To obtain the observable $N_{obs}\sim p^{-4.7}$ we should take the
source spectrum $p^{-3.85}$. The diffusion
 coefficient is rather strong function of energy $D_{\parallel }\sim p^{0.85}$.
It is interesting that such a hard spectrum of CR sources is
indeed predicted by nonlinear DSA with efficient CR acceleration
(see
 e.g. \cite{berezhko07}).

\subsection{self-consistent diffusion-advection model}

The advection of CRs in the galactic wind flow was taken into account by Ptuskin et al. \cite{ptuskin97}
In this model at small distances $s<s_*(p)$ diffusion dominates advection while at large distances the opposite
 inequality is valid. The distance to the diffusion-advection boundary $s_*(p)$ can be found from
relation $(u+v_a)s_*\sim D_{\parallel }\cos ^2\alpha $. It is possible
to use the expression for diffusion
 coefficient (10) for this estimate. It should be noted that the factor $A(s)B\cos \alpha $ is
 $s$-independent in the Galactic wind flow. This means that the
self-consistent parallel diffusion coefficient $D_{\parallel }$
does not depend on $s$.

At distances smaller than the Galactic radius $R_g$ CRs are advected with Alfv\'en velocity 
that is approximately
proportional to the height $s$. The  height
of diffusion-advection boundary is then $s_*(p)\propto p^{(\gamma -3)/2}$. The spectrum in the disk
is $N_{obs}(p)\propto Q(p)/v_a(s_*)\propto p^{-\frac 32(\gamma -1)}$.

At large distances CR particles are advected  by the wind with almost constant speed,
 the magnetic field is almost azimuthal and  $\cos \alpha \propto s^{-1}$.
This gives the distance to the diffusion-advection boundary
$s_*(p)\propto p^{(\gamma-3)/3}$. The observable spectrum
$N_{obs}\propto Q(p)/us_*^2\propto p^{-\frac 53\gamma +2}$.

We combine these cases as

\begin{equation}
N_{obs}(p)\propto
\left\{ \begin{array}{ll}
p^{-\frac 32(\gamma -1)}, \ p<p_g\\
p^{-\frac 53\gamma +2}, p>p_g
\end{array} \right.
\end{equation}
Here $p_g$ is the momentum of particles with the height of diffusion-advection boundary that is
 comparable with the Galactic radius $s_*(p_g)=R_g$. The rough estimate from Eq. (11) is $p_g\sim $ TeV/c.
So to obtain the observed spectrum $p^{-4.7}$ one should have the spectral index in the source
 $\gamma =4.13$ at $p<$ TeV/c and $\gamma =4.02$ at $p>$ TeV/c. Note that the concave spectrum of CR sources
 is indeed predicted by nonlinear DSA.

The parameter $\varepsilon $ should be adjusted to
reproduce the observed CR intensity. This gives
 the value close to $\varepsilon \approx 10^{-6}$ erg cm$^{-2}$ s$^{-1}$ \cite{ptuskin97}. The corresponding
 Galactic CR luminosity $L_{CR}=2\pi F_cR^2_g\approx 1.3\cdot 10^{41}$ erg s$^{-1}$ or $13\% $ of
 the mechanical power of Galactic supernovae. The energetic dependence and the numerical value of
 grammage $\Lambda(E)\propto E^{-0.57}$ are also in accordance with observations \cite{ptuskin97}.

Using the value of diffusion coefficient one can estimate the size
of the diffusion region $s_*(p)$. The size $s_*=1$ kpc corresponds
to the energy 10 GeV, the Galactic scale $s_*=15$ kpc corresponds
to 1 TeV. At large distances $s_*\propto p^{1/3}$ and the energy
$10^6$ GeV corresponds to the size
 $s_*=$150 kpc.

\section{Cosmic ray anisotropy problem}

Self-consistent models described above have hard spectra of
 CR sources close to $p^{-4}$. This results in fast increase of
 CR anisotropy with energy. The estimates show that the anisotropy of CR protons might
 be of the order
 of unity at PeV energies \cite{zirakashvili05}. The mixed CR composition probably will
 make the anisotropy lower close to 0.1 but still one hundred times
 higher than the observable anisotropy $10^{-3}$.

One of possibilities to avoid this contradiction is related with the small diffusion
 coefficient $D_{in}$ in the Local Bubble where the Sun is situated. The Local Bubble is a large cavity
 with the size 100 pc. It was created by OB association via
 stellar winds and supernova explosions several million years ago. Since the cavity is filled
 by the hot ionized gas the MHD waves in this region  are not damped due to presence of
 neutrals contrary to the warm ionized medium in the Bubble surroundings.

Under these conditions the anisotropy inside the Bubble is $D_{out}/D_{in}$ times smaller than
 the anisotropy outside the Bubble \cite{zirakashvili05}. Here $D_{out }$
 is the diffusion coefficient  outside the Bubble.

The main scale of turbulence can be 1 pc in the Local Bubble. The free path of CR particles
 is of the same order. Outside the Bubble the main scale and the free path of particles can be of the order
 100 pc. So the ratio of the diffusion coefficients close to one hundred is possible. This reduces
 the CR anisotropy to observable values.




\nocite{*}
\bibliographystyle{elsarticle-num}
\bibliography{martin}

\begin{thebibliography}{00}

\bibitem{ginzburg69} Ginzburg, V.L., Syrovatskii, The origin of cosmic rays, 1969,
New York : Gordon and Breach
\bibitem{berezinsky90} Berezinskii, V.S. et al., Astrophysics of
Cosmic Rays, 1990, North Holland, NY
\bibitem{krymsky77} Krymsky, G.F. A regular mechanism for the acceleration of charged
particles on the front of a shock wave, 1977, Soviet Physics-Doklady, 22, 327-329
\bibitem{bell78}Bell, A.R. The acceleration of cosmic rays in shock fronts - I, 1978, Mon. Not. Royal Astron. Soc.
 182, 147-156
\bibitem{rieger13} Rieger, F., de O\~na Wilhelmi, E., Aharonian, F., TeV astronomy, 2013,
Frontiers of Physics, 8, 714-747
\bibitem{obermeier12} Obermeier, A., Boyle, R., H\"orandel, J., M\"uller, D.,
The Boron-to-carbon Abundance Ratio and Galactic Propagation of Cosmic Radiation, 2012,
Astrophys. J., 752, 69-75
\bibitem{brandenburg13} Brandenburg, A., Lazarian, A., Astrophysical Hydromagnetic Turbulence, 2013,
Space Science Reviews, 178, 163-200
\bibitem{strong07} Strong, A.W., Moskalenko, I.V., Ptuskin, V.S. 
Cosmic-Ray Propagation and Interactions in the Galaxy, 2007, 
Annual Review of Nuclear and Particle Systems 57, 285-327
\bibitem{yan04} Yan, H., Lazarian, A., Cosmic-Ray Scattering and Streaming in Compressible
Magnetohydrodynamic Turbulence, 2004, Astrophys. J. 614, 757-769
\bibitem{ginzburg61} Ginzburg, V.L., Propagation of Electromagnetic Waves in Plasma, 1961, New York:
Gordon \& Beach
\bibitem{farmer04} Farmer, J.F., Goldreich, P., Wave Damping by Magnetohydrodynamic
Turbulence and
Its Effect on Cosmic-Ray Propagation in the Interstellar Medium, 2004, Astrophys. J. 604, 671-674
\bibitem{kulsrud69} Kulsrud, R.M., Pearce, W.P.
The effect of wave-particle interactions on the propagation of cosmic rays, 1969, Astrophys. J.
156, 445-470
\bibitem{wentzel69} Wentzel, D.G. The Propagation and anisotropy of cosmic rays. I.
Theory for steady streaming, 1969, Astrophys. J., 156, 303-314
\bibitem{kulsrud71} Kulsrud, R.M., Cesarsky, C.J.
The effectiveness of instabilities for the confinement of high energy cosmic rays in the Galactic disk, 1971,
Astrophys. Lett. 8, 189
\bibitem{ginzburg73} Ginzburg, V.L., Ptuskin, V.S., Tsytovich, V.N. The role of plasma effects in
propagation and isotropisation of cosmic rays in the Galaxy, 1973, Astrophys. Space Sci. 21, 13-38
\bibitem{skilling75} Skilling, J.
Cosmic ray streaming. III - Self-consistent solutions, 1975,
Mon. Not. Royal Astron. Soc. 173, 255-269
\bibitem{lerche67} Lerche, I. Unstable magnetosonic waves
in a relativistic plasma, 1967, Astropys. J. 147, 689-696
\bibitem{lee73}
Lee, M.A., V\"{o}lk, H.J. Damping and non-linear wave-particle
interactions of alfv\'en-waves in the solar wind, 1973, Astrophys. and Space
Sci. 24, 31-42
\bibitem{achterberg81}
Achterberg, A. On the propagation of relativistic particles in a
high beta plasma, 1981, Astron. Astrophys. 98, 161-172
\bibitem{kulsrud82}
Kulsrud, R.M. Plasma in astrophysics, 1982,  Physica Scripta 2/1, 177-181
\bibitem{achterberg86}Achterberg, A., Blandford, R.D. Transmission and damping of hydromagnetic waves
behind a strong shock front: implications for cosmic ray acceleration, 1986,  Mon. Not. Royal Astron. Soc.
218, 551-575
\bibitem{fedorenko88}
Fedorenko, V.N., Ostryakov, V.M., Polyudov, A.N., Shapiro, V.D.
Induced scattering and two quantum absorption of Alfv\'en waves in plasma with
arbitrary beta, 1988, Preprint N 1267 A.F.Ioffe Phys.Tech. Inst., Leningrad
\bibitem{zirakashvili00} Zirakashvili, V.N., Induced Scattering and
Two-Photon Absorption of Alfven Waves with Arbitrary Propagation Angles, 2000, JETP 90, 810-816
\bibitem{ipavich75}
Ipavich, F.M. Galactic winds driven by cosmic rays, 1975, Astrophys. J. 196, 107-120
\bibitem{breitschwerdt91}
Breitschwerdt, D., McKenzie, J.F., V\"{o}lk, H.J. Galactic Winds. I
- Cosmic ray and wave-driven winds from the Galaxy, 1991, Astron. 
Astrophys. 245, 79-98
\bibitem{zirakashvili96}
Zirakashvili, V.N., Breitschwerdt, D., Ptuskin, V.S., V\"olk, H.J.
Magnetohydrodynamical galactic wind driven by cosmic rays in a
rotating galaxy, 1996,  Astron. Astrophys. 311, 113-126
\bibitem{everett07} Everett, J.E., Zweibel, E.G.; Benjamin,R.A., et al. 
Does the Milky Way launch a large-scale wind? 2007, Astrophys. and Space Science, 311, 105-110
\bibitem{miller13} Miller, M.J., Bregman, J.N., The Structure of the Milky Way's
Hot Gas Halo, 2013, Astrophys. J., 770, 118-130
\bibitem{ptuskin97}Ptuskin, V.S., Zirakashvili, V.N., Breitschwerdt, D., V\"olk, H.J.
Transport of relativistic nucleons
in a galactic wind driven by cosmic rays, 1997, Astron. Astrophys. 321, 434-443
\bibitem{zirakashvili02} Zirakashvili, V.N., Ptuskin, V.S., V\"olk, H.J. 2002,
Bull. Russian Ac. Sci. 66, 1606-1608 (in Russian)
\bibitem{alloisio13} Alloisio, R., Blasi, P., Propagation of galactic cosmic rays in the
presence of self-generated turbulence, 2013, 
Journal of Cosmology and Astroparticle Physics, 7, 1-23 
\bibitem{berezhko07} Berezhko, E.G., V\"olk, H.J., Spectrum of Cosmic Rays Produced in Supernova Remnants, 2007
Astrophys. J. 661, L175-L178
\bibitem{zirakashvili05} Zirakashvili, V.N. Cosmic ray anisotropy problem, 2005,
Intern. Journal of Modern Physics A 20, 6858-6860

\end{thebibliography}



\end{document}